\documentclass[
   ,final            
  ]
  {aipproc}

\layoutstyle{6x9}

\usepackage{amsmath, amssymb}
\usepackage{pifont}

\begin{document}

\title{Cloud formation in substellar atmospheres}

\classification{95.30.Wi, 97.10.Ex, 97.20.Vs}
\keywords      {astrochemistry, stars: atmospheres, stars: low-mass, brown dwarfs, infrared: stars  }

\author{Christiane Helling}{
  address={SUPA, School of Physics and Astronomy, Univ. of St Andrews, North Haugh,\\ St Andrews,  KY16 9SS, UK}
}

\begin{abstract}
Clouds seem like an every-day experience. But -- do we know how clouds
form on brown dwarfs and extra-solar planets? How do they look like?
Can we see them?  What are they composed of?  Cloud formation is an
old-fashioned but still outstanding problem for the Earth atmosphere,
and it has turned into a challenge for the modelling of brown dwarf
and exo-planetary atmospheres.  Cloud formation imposes strong
feedbacks on the atmospheric structure, not only due to the clouds own
opacity, but also due to the depletion of the gas phase, possibly
leaving behind a dynamic and still supersaturated atmosphere.
I summarise the different approaches taken to model cloud formation in
substellar atmospheres and workout their differences. Focusing on the
phase-non-equilibrium approach to cloud formation, I demonstrate the
inside we gain from detailed micro-physical modelling on for instance
the material composition and grain size distribution inside the cloud
layer on a Brown Dwarf atmosphere.  A comparison study on four
different cloud approaches in Brown Dwarf atmosphere simulations
demonstrates possible uncertainties in interpretation of observational
data.

\end{abstract}

\maketitle


\section{Introduction}
Brown Dwarfs are object much more massive than planets, but they can
become as cold as planets during their lifetime. Young gas giants and
also so-called close-in giant gas planets can be as hot as Brown
Dwarfs (e.g. \cite{ca2008}, \cite{heb2008}).
Brown Dwarfs can only nucleosynthesis energy during a short period of
their lift time, which is enough to bust their luminosity to a level
where they can be much easier observed than any planet at a similar
distance. The largest similarity between Brown Dwarfs and planets is
the rich, multi-phase atmospheric chemistry.  Descending the main
sequence makes Brown Dwarfs and possibly late-type M-dwarfs the only
objects, that form like stars and that contain dust (i.e. small solid
particles) in their atmosphere. After this discovery
\cite{Jones1997}, it became immediately clear that Brown Dwarfs were
not the perfect example for a static atmosphere but feedback of the
dust clouds on the atmospheric structure, the chemistry, and the
radiative transfer would need attention. Other places of efficient
dust formation are the circulstellar envelopes of AGB stars which are
extremely dynamic systems due to the presence of dust (see review by
A.C. Andersen, and contributions by e.g. S.H\"ofner, N. Thureau this
volume). \cite{cu2008}\, have presented spectra for L-type Brown
Dwarfs showing broad absorption features between 8$\mu$m and 11$\mu$m
which is the wavelength range were silicate dust would
absorb. Although the idea of dust absorbing in the atmosphere is
thermodynamically correct, it is hard to proof its existence directly
since we are basically searching for the absorption signature of a
thin transparent layer on top of an optically thick wall
(\cite{hel2006}). \cite{gel2002}\, concluded that inhomogeneities in
cloud decks and the evolution thereof can plausibly produce observed
photometric $I$-band variations.  Extensive studies searching for
uncorrelated, time-dependent variability due to a variable cloud
coverage of the atmosphere, suggest a variability of 2--3\%
(\cite{gold2008}) or even 2--10\% (\cite{bj2008}) depending in the
wavelength studied.  Richardson et al. (2007) infer the presence of
silicate haze based on secondary-eclipse Splitzer observations on
giant gas-planets, a conclusion which was also put forward by
\cite{po2008} based on HST transmission spectra for HD\,189733b (see
also reviews by G. Tinetti and by J. Harrington, this volume).

\begin{figure}
  \includegraphics[height=.3\textheight]{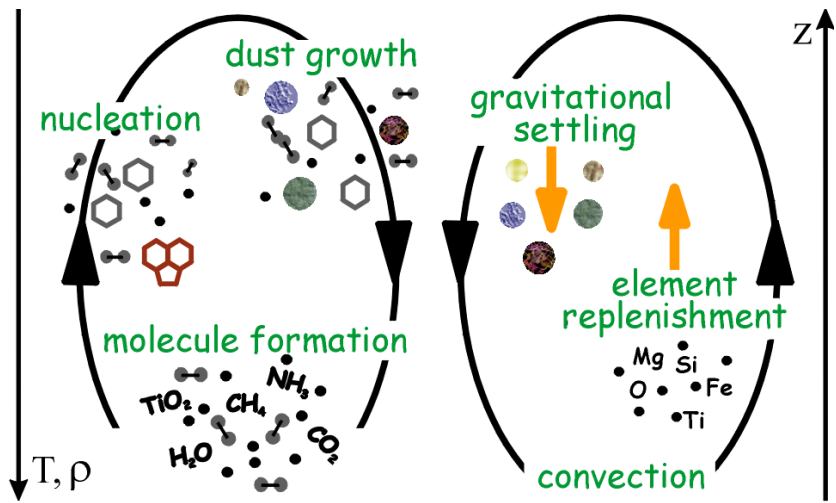}
  \caption{Interaction of multi-scale processes during cloud formation
  in Brown Dwarf atmospheres.Dust formation and turbulence act on
  small scales, and gravitational settling and element replenishment
  act on large scales. }
\label{f:circ}
\end{figure}

\section{Cloud formation processes}\label{s:cp}

Three major complexes need to be modelled to account for cloud
formation in model atmospheres and to produce observable
informations. Figure~\ref{f:circ} demonstrates how condensation,
gravitational settling and element replenishment interact.

\begin{itemize}
\item{\sf Condensation:}\\
Seed particle form from the gas phase via a net or a path of chemical
reaction. Once the first solid surface appears from the gas phase,
chemical surface reaction can proceed rapidly to grow a massive mantle
such that the initially tiny seed growth to a macroscopic grain of
$\mu$m-size. While the nucleation process is only possible in a highly
supersaturated gas, the growth can take place if a solid is thermally
stable. The counter-acting process is evaporation if the temperature
becomes too high. The condensation process reduces the element
abundances is the gas phase selectively regarding those element
forming grain monomers (like MgSiO$_3$, TiO$_2$, SiO$_2$ in the
MgSiO$_3$[s]-, TiO$_2$[s]-, SiO$_2$[s]-lattice etc.).
\item{\sf Gravitational settling}\\
Once the grains are present they start to fall into the atmosphere due
to the strong gravitational attraction in Brown Dwarfs and in
planets. Since the grain sizes are not constant but different grain
sizes will appear with different grain numbers, the equation of motion
would need to be solved for a grain size distribution. An easier case
is to solve the equation of motion for a height-dependent mean grain
size derived from a grain size distribution.
\item{\sf Element replenishment}\\
A static atmosphere would be dust free (\cite{wo2004}).
\cite{op1998}, \cite{sau2007} and \cite{le2007} infer from their observations that convection mixes up
(elementally rich) gas from inner and hotter atmospheric regions, a
mechanism essential for allowing the dust cloud to persist inside the
atmospheres.  The friction of the gas fluid is not very large so that
its inertia drives it forward once it has been set into motion. Such
fluids are called {\it turbulent} where whirly motions exist on a
large range of scale. Convection is just one of them, but the largest
on which the turbulent fluid field is energetically generated. Hence,
modelling element replenishment means to describe mixing, advection,
overshooting, and turbulence.
\end{itemize}

\noindent
As soon as the first Brown Dwarf was discovered, model atmosphere
predictions where needed, and three different ideas appeared regarding
the phenomenology of cloud formation: \\[-0.7cm]
\begin{itemize}
\item[i)] {\sf Static:}  The cloud particle are so small that they remain where they formed. \\(Tsuji, Copper et al.),
\item[ii)] {\sf Bottom-up:} The cloud forms by  up-mixing of uncondensed material. \\(Ackerman \& Marley, Allard et al.)
\item[ iii)]  {\sf Top-down:} The cloud is formed by particles falling into the atmosphere and evolving on their way downward. (Helling \& Woitke)
\end{itemize}
\noindent
 Table ~\ref{t:modellers} summarises the cloud models and their
variants presently applied in the literature. The grain size is an
opacity input quantity, and the very first models set it constant
assuming the grains to hoover inside the atmosphere (Tsuji). Later,
experiences from cloud observations on Earth led to a time-scale
comparisons to derive the grain sizes (Allard et al.) or to an
analytic parameterisation thereof (Cooper et al., Ackerman \&
Marley). Only very recently, a grain size distribution based on a
model for dust formation was derived (Helling
\& Woitke).  The calculation of the dust cloud opacity needs further
the material composition of the cloud. The modelling of this is very
much linked to the model assumption regarding the formation
process. Almost all models assume phase-equilibrium between the
atmospheric gas and the cloud particles, hence the supersaturation
ratio $S=1$, or just a 1\% supersaturation.  However, only the departure from
this equilibrium state, i.e. $S\gg 1$, would allow the actual grain
formation. Assuming phase-equilibrium allows to consider the
thermal stability of a vast variety of potential condensates
(e.g. \cite{lo2003}) which then can be considered to deplete the
gas-phase.  This phase-equilibrium approach bears the danger that one
considers compounds which might never have a chance to actually form.

\begin{table}
\resizebox{10cm}{4cm}{
\hspace*{-2.5cm}
\begin{tabular}{l|l|l|l|ll}
                     & \multicolumn{3}{c|}{{\bf Assumptions}}& \multicolumn{2}{c}{\ } \\
{\bf Author}  &  grain size $a$ & grain       & super- &    \multicolumn{2}{c}{\bf Model variants} \\
           &                 & composition & saturation &  \\
\hline
Tsuji          & $a=10^{-2}\mu$m & homog. & $S=1$ & {\it case B}       & full dusty model \\
(also Burrows 2001 )               &                 &             &      & {\it case C}  & dust cleared model \\
               &                 &             &         & {\it UCM}     & dust between\\
               &                 &             &       &               &  $T_{\rm cr}<T<T_{\rm cond}$ \\
               &                 &             &         &               & \cite{ts2000}, \cite{ts1996}, \cite{ts2002}, \cite{ts2005}\\[0.2cm]
 Allard \& Homeier& $f(a)=a^{-3.5}$    & homog. & $S=1$ &{\it dusty} & full dusty model\\
               &          &            &        &{\it cond}          & dust cleared model\\
               &time scales dep.   & homog. & $S=1.001$ &{\it settl} & time scales \\
               &          &            &        &       &            \cite{all2001}, \cite{all2003}, \cite{all2007}, \cite{ro1978}\\ [0.2cm]
 Copper et al. &  $f(a)\sim\big(\frac{a}{a_0}\big)^6$ & homog. &$S=1.001$ &  &\cite{co2003} \\
 & $\times\exp\big[-6\big(\frac{a}{a_0}\big)\big] $ & & & & \\[0.2cm]
Ackerman \& Marley  & log-normal $f(a,z)$ & homog. & $S=1$ & $f_{\rm sed}$ & sedimentation\\
    \& Lodders                              & &  &      &    &              efficiency \cite{ack2001}\\[0.2cm]
Helling \& Woitke   & $f(a, z)$           & dirty  & $S=S(z,s)$&&\cite{wo2003}, \cite{hel2008a}
\end{tabular}
\caption{Cloud models in Brown Dwarf atmosphere simulations.
}
\label{t:modellers}
}
\end{table}

Despite the variety of the assumptions made to model clouds in Brown
Dwarf atmospheres (Table~\ref{t:modellers}), fundamental understanding
has been added by considering the limiting cases {\it case B / dusty}
and {\it case C / cond} by Tsuji and Allard et al., respectively. The
dust is element sink and opacity source inside the atmospheres in {\it
case B / dusty} which represents the the mid-L-dwarf regime. {\it Case
C / cond} do consider the dust as element sink only and neglect its
opacity. This case represented the T-dwarfs regime very well. Hence,
these two simple limiting cases have established our picture from
L-dwarfs as being covered in thick clouds, and of the T-dwarfs as
Brown Dwarfs with an elementally depleted atmosphere but without
visible clouds. As question remains how the clouds disappear going
from L- to T-dwarfs (L\,-T\, transition region, \cite{kna2004},
\cite{gol2004}). The main idea is that the cloud thickness
changes. This is parameterised by a critical temperature $T_{\rm cr}$
(\cite{ts2002}), or a sedimentation efficiency $f_{\rm sed}$
(\cite{ack2001},\cite{cu2008}), or the mixing efficiency (Allard et
al.). As long as no consistent theory exists to allow the simulation
of the L\,-T\,, such empirical parameterisations are needed and
modellers try to constrain them from observations (\cite{ts2005};
Stephens et al., this volume). This inconsistency complicates also the
evolutionary modelling from the L- into the T-dwarf regime as
discussed in
\cite{sau2008}.

\section{The phase-non-equilibrium approach}

The cloud {\it formation} process can only proceed in non-equilibrium
state, else the system would not change at all. A phase-transition
will therefore require a phase-non-equilibrium, hence, the gas phase
needs to be supersaturated ($S\gg 1$). Depending on time and grain
size scales, the dust formation proceeds via 1.) nucleation, followed
by 2.) growth (Sect.~\ref{s:cp}).  Nucleation runs on much shorter
time scales than the growth process and needs a highly supersaturated
gas to allow a reaction efficiency high enough to counter-balance
destructive backward reactions. The (not-yet) particles which
subsequently form in the nucleation regime are macro-molecules and
clusters (e.g. \cite{goe1993, pa1998}).  The result is a nucleation
rate which determines the number of dust particles. \cite{jeo96,
jeo00} have shown that TiO$_2$ is a suitable candidate for seed
formation in oxygen-rich environments since enough molecular TiO$_2$
is available to allow an efficient homogeneous cluster formation
process, in fact much more efficient than for instance SiO- or
Fe-nucleation. Once, such a first surface emerges from the gas phase,
lots of other compounds are already thermally stable and can therefore
grow by chemical surface reactions. Still, thermal stability is not
the only criterion, but we need to know which chemical surface
reactions are possible (see
\cite{hel2008a}). A straight forward observation from
gas-phase chemical equilibrium calculations is, that only small
molecules like MgO, MgOH, SiO, SiS, Al$_2$O, Ca(OH)$_2$ are abundant
enough to allow high growth rates. This conclusion is supported by
condensation experiments \cite{rid1999, rid2007}.

\begin{figure}
  \includegraphics[height=.26\textheight]{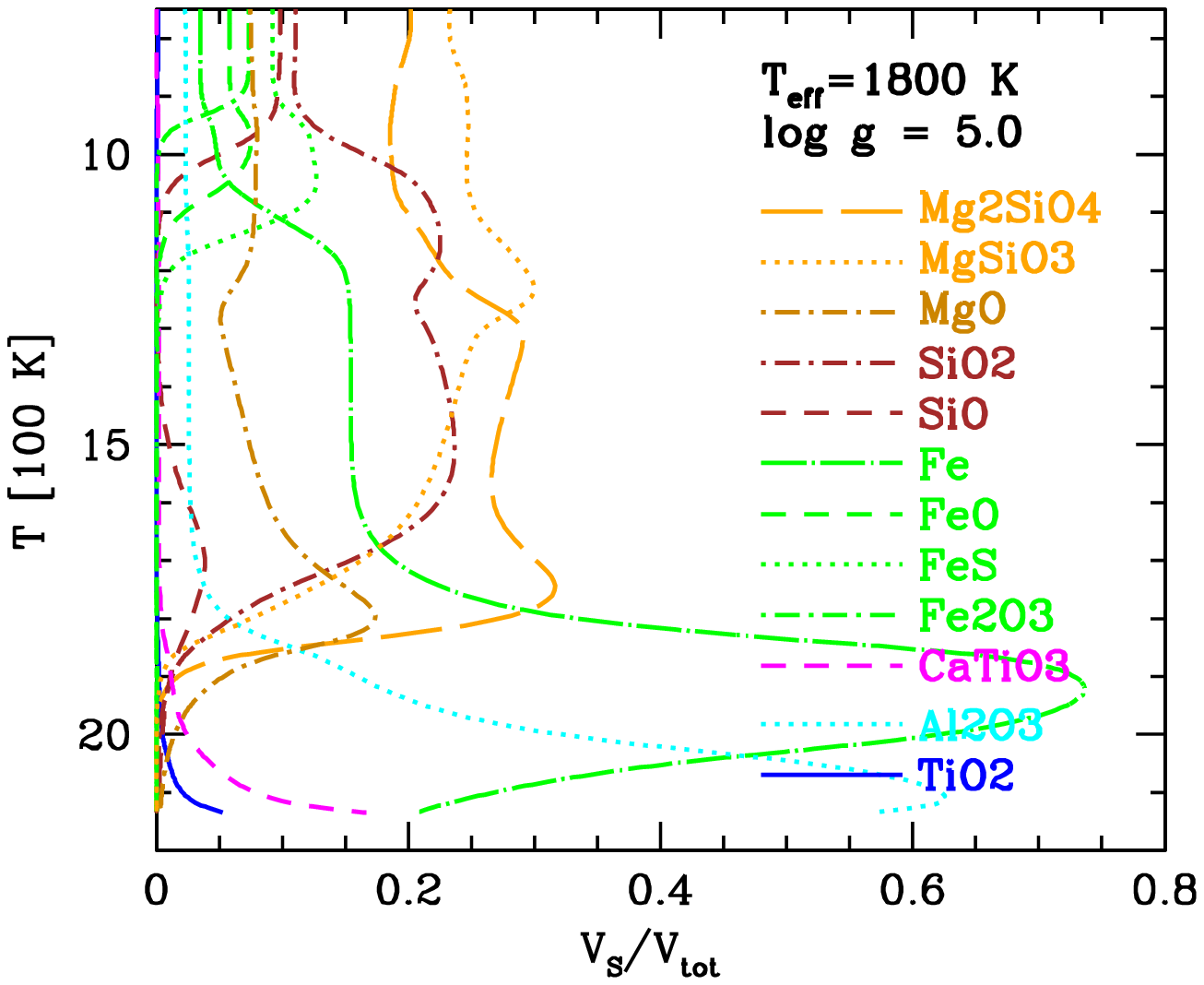}
  \includegraphics[height=.26\textheight]{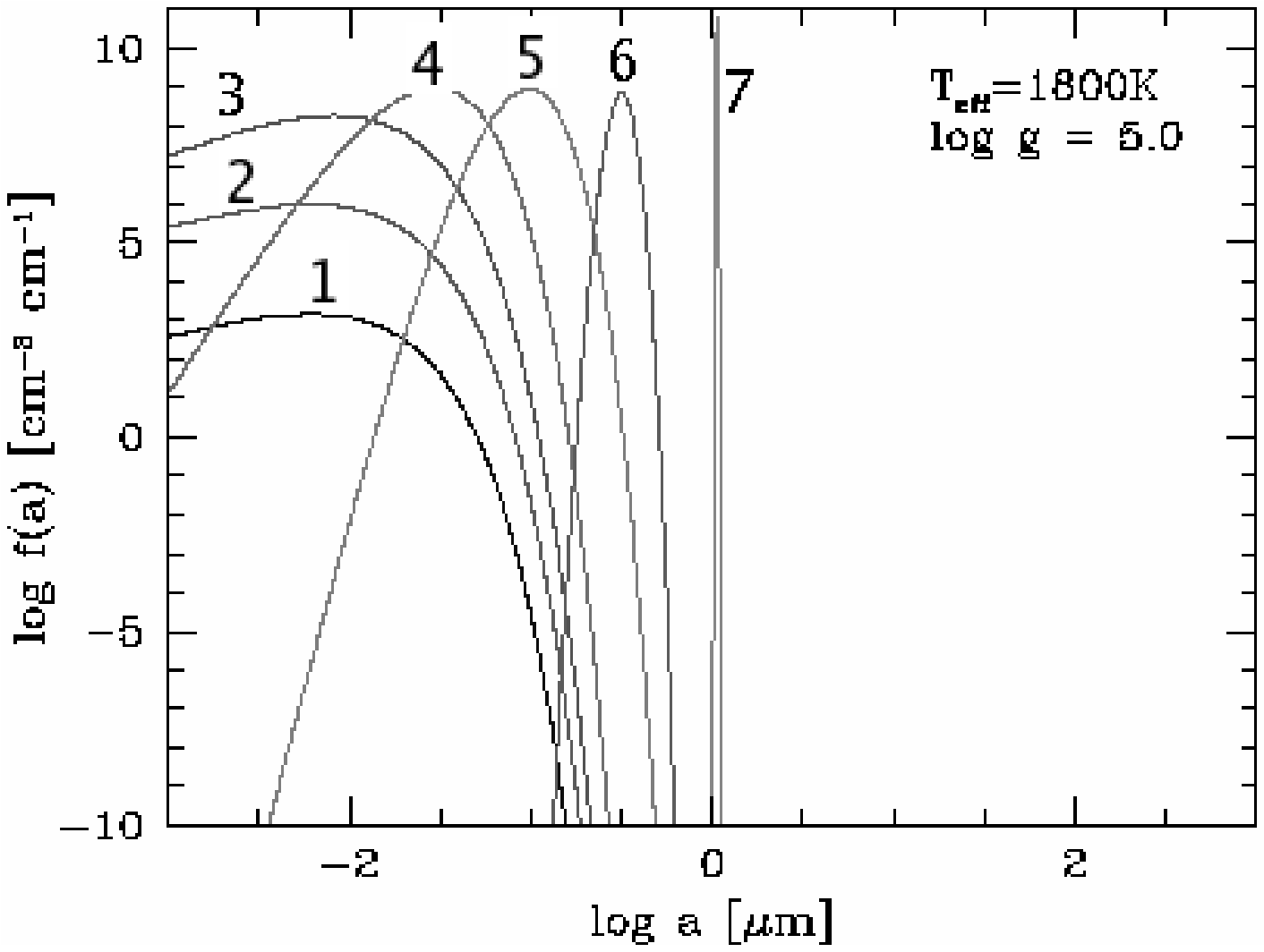}
  \caption{Material composition (in volume fractions $V_{\rm s}/V_{\rm
  tot}$; left) and altitude-dependent grain size distributions ($f(a)$
  in cm$^{-3}$cm$^{-1}$; right) in a stationary dust cloud layer in a
  Brown Dwarf atmospheres. }
\label{f:Vol_Fa}
\end{figure}

The treatment of TiO$_2$-nucleation by a modified classical nucleation
theory \cite{jeo00}, and the calculation of the grow process including
the effect of drift in connection with a parametrised up-mixing of
uncondensed gas (\cite{wo2003, wo2004, hel2006, hel2008a}) allows to
study the cloud's material composition, the altitude-dependent grain
size distribution, and also the phase-equilibrium state of the
dust-gas mixture in the cloud beside various other details of the dust
complex like nucleation rate, growth velocity, dust-to-gas ratio,
metallicity. Figure~\ref{f:Vol_Fa} (left) shows that the cloud deck is
populated by silicate grains mainly made of MgSiO$_3$[s] and
Mg$_2$SiO$_4$[s] beside some iron-compounds. The MgSiO$_3$[s] drops
with increasing temperate which increases the fraction of SiO$_2$[s]
inside $\approx 1000$K. All silicates are evaporated at $T\approx
2000$K, and the cloud particles are now composed of mainly Fe[s], and
Al$_2$O$_3$[s] at even higher temperatures. Figure~\ref{f:Vol_Fa}
(right) shows a number of selected grain size distribution function,
$f(a)$ [cm$^{-3}$ cm$^{-1}$], sampling the cloud layer of that
model. The lowest curve (1) represent the outermost grain
size distribution in the region of nucleation of those depicted
here. All subsequent $f(a)$ (2 -- 7) move towards larger grain sizes
(abscissa) but only numbers 1, 2, 3 show an increase in numbers of
grains. 4, 5, 6 have achieved the maximum possible number of grains
and only move in grain-space to the right, hence, they represent cloud
layers with continuously increasing grain sizes.  4, 5, 6 also
narrower which demonstrates that the cloud base is populates by a
relatively narrow interval of grain sizes. These big grains are mainly
made of Fe[s], or Al$_2$O$_3$[s] just before they evaporate.

\begin{figure}
  \includegraphics[height=.4\textheight]{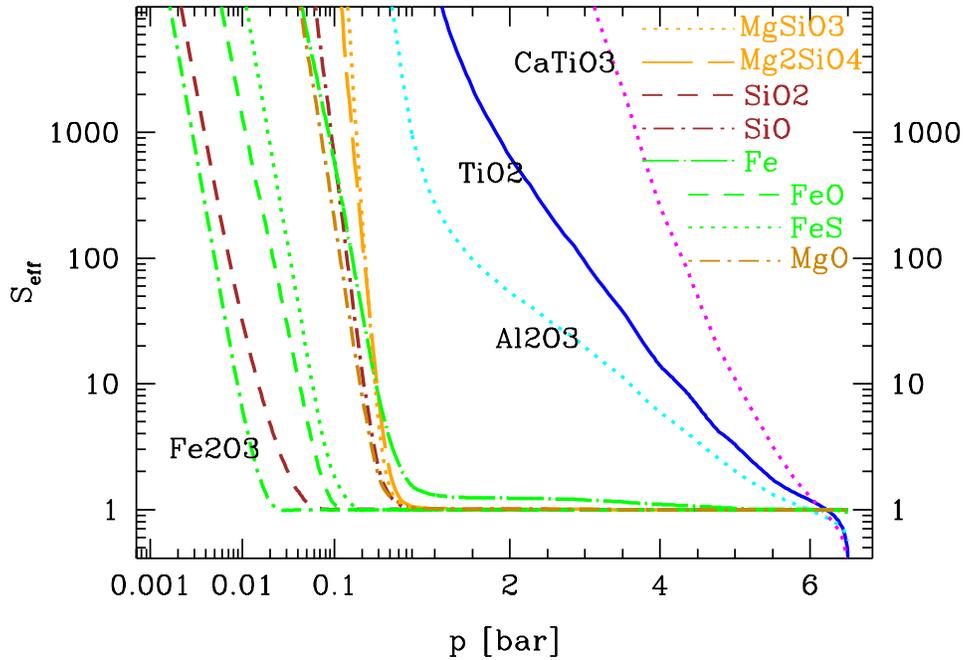}
  \caption{Effective supersaturation ratio ($S_{\rm eff}$) of the dust
  compounds considered in Fig.~\ref{f:Vol_Fa}. Note that $S_{\rm eff}$
  has been defined in (Helling, Woitke \& Thi 2008) to discuss the
  contribution of all growth reaction of a particular compound
  compared to the standard definition.}
\label{f:Seff}
\end{figure}

The phase-non-equilibrium approach discussed so far does not assume a
certain supersaturation ratio, S, but calculates it according to the
present thermodynamical conditions and element abundances. We are
therefore in the position to study if and where the dust-gas mixture
forming the cloud layer is in phase equilibrium or not. Hence, we can
study where the cloud is thermally stable ($S=1$) and where is is
effectively processed in a constructive ($S>1$) or destructive ($S<1$)
way. Figure~\ref{f:Seff} depicts the effective supersaturation ratio,
$S_{\rm eff}$\footnote{The classical supersaturation ratio assumes
that the compound monomer (like a Mg$_2$SiO$_4$-molecule for
Mg$_2$SiO$_4$[s]-solid) does exist in the gas phase. Since this is not
always the case, the supersaturation ratio for every surface reaction
needs to be calculated since here the gas-phase constituents are
known. For further details, see \cite{hel2006}.}, defined to
take into account the contribution of all growth reaction for a
particular compound. The compounds considered fall into two regimes:
\begin{itemize}
\item {\sf rare-element compounds:}\\
 Ti-, Al-, Ca-solids exhibit an extreme phase-non-equilibrium
\item {\sf abundant-element compounds:}\\ 
Mg-, Si-, Fe-solids achieve phase-equilibrium over some pressure scale height in the cloud once the seed particles serve as condensation surface
\end{itemize}

\noindent
Figure~\ref{f:Seff} shows that also Fe[s] remains in
phase-non-equilibrium in almost the entire cloud layer, hence only the
low-temperature condensates can achieve
phase-equilibrium. Consequently, phase-non-equilibrium models would
show larger gas-phase abundances regarding molecules containing
element of thermally very stable compounds as demonstrated in
\cite{hel2008b}.

\begin{table}
\resizebox{14cm}{!}{
\begin{tabular}{ccccccl}
{\bf authors} & {\bf element}    & {\bf  elements} & {\bf gas-phase}       & {\bf number of}    \\
                       & {\bf abund.}     &                            & {\bf spec.} & {\bf dust species} \\
\hline
{\bf Tsuji}             & \cite{and1986} & 34 & 83 & 3 as opacity source   & s\\
                        & \cite{alle2002} &    &      & 10 as element sinks & s\\[0.3cm]
{\bf Allard \& Homeier}  &\cite{grev1992}     & 84 & 680 & 43 as opacity source & sl\\
                        & \cite{asp2005}&    &    &     169 as element sinks   & sl\\[0.3cm] 
{\bf Marley, Ackerman}  & \cite{lo2003}  & 83 & $\sim 2200$  & 5 as opacity source & s\\
{\bf \& Lodders}                         &      &    &    & $\sim 1700$ as element sinks & sl\\[0.3cm]
{\bf Dehn \& Hauschildt}   & \cite{grev1992} & 40 & 338 & 7 as opacity source & s  \\
{\bf + Helling \& Woitke} &   &    &     & 7 as element sinks & s\\[0.5cm]
\end{tabular}
\caption{Brown Dwarf model atmosphere codes in comparison. See also Table~\ref{t:modellers}.\newline s -- solid, sl -- solid \& liquid.}
}
\label{t:part}
\end{table}

\begin{figure}
  \includegraphics[height=.6\textheight]{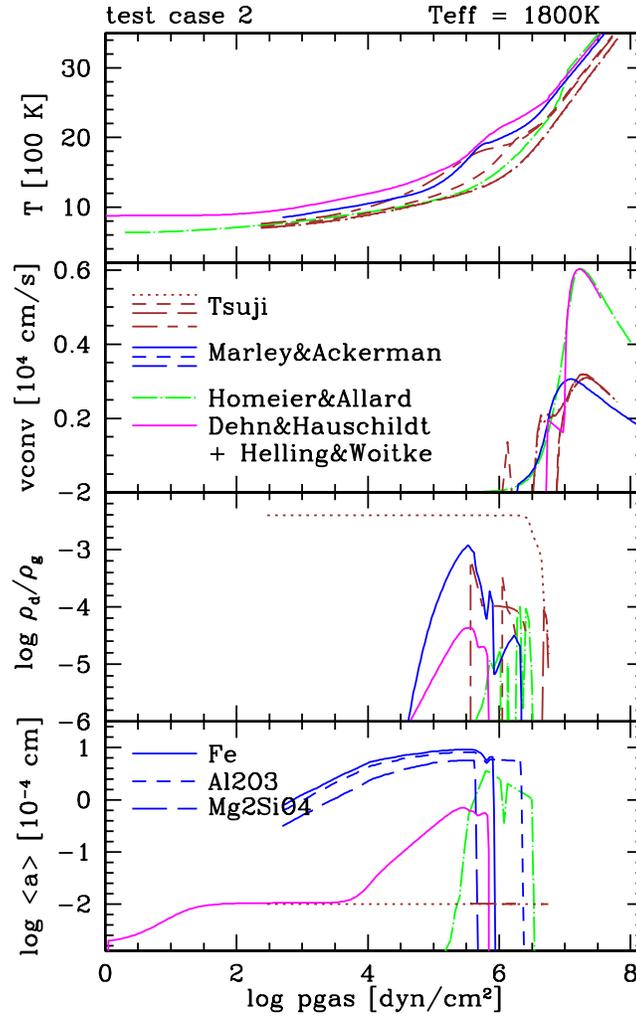}
  \caption{Test cases for complete atmospheric models for $\log$\,g =
  5.0, solar element abundance with T$_{\rm eff}=1800$K.  {\bf Note:}
  Different colours stand for different stellar atmospheres
  codes. Four models are plotted for the Tsuji-case (brown):
  long-short-dashed: T$_{\rm cr}=$1700K (extended cloud),
  short-dashed: T$_{\rm cr}=$1800K, long-dashed: T$_{\rm cr}=$1900K
  (thin cloud), dotted: no dust opacity considered.  Different line
  styles in $\log \langle a \rangle$ indicate different homogeneous
  dust species in the Marley, Ackerman \& Lodders-models.}
\label{f:compall}
\end{figure}

\section{Comparing dust cloud models}
{\it ``Where stability and equilibrium are inherent signs if an unchanging
and 'static' world, the formation and evolution of complex physical
systems is intimately coupled to the existence and stabilisation of
non-equilibrium states of matter.''} \cite{pa1998}. 

 Section~\ref{s:cp} already hints that present cloud models are well
 described by the above statement taken from  \cite{pa1998}. Namely,
 one kind of approaches studies the possible end-state of the cloud
 formation processes which is the (phase-)equilibrium state, the other
 kind has its emphasis on the formation process which is the
 (phase-)non-equilibrium state. Both aim to understand the influence
 of atmospheric dust clouds on observable quantities by means of model
 atmospheres simulations.  A comparison study of such models with
 emphasis on the dust approaches was initiated at a workshop in Leiden
 2006(\footnote{http://www.lorentzcenter.nl/lc/web/2006/203/info.php3?wsid=203}). Table~\ref{t:part}
 demonstrates that not only the cloud models are different
 (Table~\ref{t:modellers}) but also the way the results of the cloud
 model approaches are used in the radiative equilibrium calculation
 differ.  The number of compounds considered as opacity species is
 usually smaller than the number of compounds used as element sinks,
 and the element abundance data are taken from different sources. Part
 of the reason are missing refractory index data and numerical
 difficulties in handling such complex absorbing systems. The element
 abundances are particular interesting since they determine the
 composition of the gas phase (beside the local temperature and
 density) from which the dust forms.

 Figure~\ref{f:compall} compares the results from four different model
 atmospheres codes (Table~\ref{t:part}) for the stellar parameter
 combination of a typical solar-metallicity L-dwarf. The
 temperature-pressure structures, $(T,p)$ (1$^{\rm st}$ panel), differ
 the most in the low-pressure regime above the cloud layer and in the
 $(T,p)$-range where the cloud sits. The models which produce a thick
 cloud layer show the backwarming effect due to the strong cloud
 opacity at $\approx 2000$K.  The dust-to-gas mass ratio, $\rho_{\rm
 d}/\rho_{\rm g}$ (3$^{\rm rd}$ panel), demonstrates where most of the
 dust mass is located, and coincides with the backwarming
 effect. However, the $\rho_{\rm d}/\rho_{\rm g}$ maxima differ widely
 amongst the models. Such a strong difference must be attributed to
 the amplification of small differences in a coupled system such as a
 stellar atmosphere, because Helling et al. (2008) demonstrate much
 smaller differences in the $\rho_{\rm d}/\rho_{\rm g}$ maxima if the
 dust models alone are tested.  The models also differ in the
 predicted mean grain sizes, $\langle a\rangle$ (4$^{\rm th}$ panel),
 which reflects the differences in the cloud model approaches the
 strongest amongst the quantities in Fig.~\ref{f:compall}.

\section{Conclusions}

The modelling of Brown Dwarf atmospheres has turned into an unexpected
challenge as these atmosphere contain dust which causes strong
feedbacks as opacity source and element sink. Different approaches are
applied to model dust in Brown Dwarf atmospheres, and obeying its
nature as coupled system of nonlinear equation, atmosphere simulations
applying different cloud models do produce different results
although the over-all cloud-structure appears robust. A fundamental
difference amongst the models is the consideration of
phase-equilibrium. Kinetic models did show that in particular
compounds made or rare elements like Ti, Al, and Ca never achieve
phase equilibrium, i.e. thermal stability, in the Brown Dwarf
atmosphere. Furthermore, the transition from the L- into the T-dwarf
regime is not understood in enough detail that consistent atmosphere
simulations are available. This is a sever challenge for evolutionary
models.

What else needs to be done? A still outstanding problem is the
modelling of turbulent dust formation in large scale simulations which
is also of significance for planetary atmospheres or protoplanetary
disks. The challenge is here not to lose the small-scale interactions
between the chemistry and the fluid field which has been shown to
support the turbulent nature of the fluid field and to even cause the
appearance of medium-scale cloud structures. A second challenge is the
possible coupling of the dust-ionised atmosphere to the object's
magnetic field. \cite{mo2002} and \cite{gel2002} argue
against an ionisation potential in Brown Dwarf atmospheres. However,
drifting grains in a turbulent environment might make a
re-consideration necessary.





\bibliographystyle{aipproc}   


\IfFileExists{\jobname.bbl}{}
 {\typeout{}
  \typeout{******************************************}
  \typeout{** Please run "bibtex \jobname" to optain}
  \typeout{** the bibliography and then re-run LaTeX}
  \typeout{** twice to fix the references!}
  \typeout{******************************************}
  \typeout{}
 }

\end{document}